\documentclass[twocolumn,aps,prd,superscriptaddress,showpacs,nofootinbib,floatfix,letterpaper]{revtex4}

\usepackage{amsmath}
\usepackage{graphicx}   
\usepackage{xspace}     
\usepackage{verbatim}   

\newcommand{\pizero}{\ensuremath{\pi^0}\xspace}

\begin{document}

\title{Measurement of transverse-single-spin asymmetries for midrapidity 
  and forward-rapidity production of hadrons in polarized $p$$+$$p$
  collisions at $\sqrt{s}=200$ and 62.4~GeV}

\newcommand{\abilene}{Abilene Christian University, Abilene, Texas 79699, USA}
\newcommand{\acadsin}{Institute of Physics, Academia Sinica, Taipei 11529, Taiwan}
\newcommand{\augie}{Department of Physics, Augustana College, Sioux Falls, South Dakota 57197, USA}
\newcommand{\banaras}{Department of Physics, Banaras Hindu University, Varanasi 221005, India}
\newcommand{\barc}{Bhabha Atomic Research Centre, Bombay 400 085, India}
\newcommand{\baruch}{Baruch College, City University of New York, New York, New York, 10010 USA}
\newcommand{\bnlcoll}{Collider-Accelerator Department, Brookhaven National Laboratory, Upton, New York 11973-5000, USA}
\newcommand{\bnlphys}{Physics Department, Brookhaven National Laboratory, Upton, New York 11973-5000, USA}
\newcommand{\caucr}{University of California - Riverside, Riverside, California 92521, USA}
\newcommand{\charlesczech}{Charles University, Ovocn\'{y} trh 5, Praha 1, 116 36, Prague, Czech Republic}
\newcommand{\chonbuk}{Chonbuk National University, Jeonju, 561-756, Korea}
\newcommand{\ciae}{Science and Technology on Nuclear Data Laboratory, China Institute of Atomic Energy, Beijing 102413, P.~R.~China}
\newcommand{\cns}{Center for Nuclear Study, Graduate School of Science, University of Tokyo, 7-3-1 Hongo, Bunkyo, Tokyo 113-0033, Japan}
\newcommand{\colorado}{University of Colorado, Boulder, Colorado 80309, USA}
\newcommand{\columbia}{Columbia University, New York, New York 10027 and Nevis Laboratories, Irvington, New York 10533, USA}
\newcommand{\czechtech}{Czech Technical University, Zikova 4, 166 36 Prague 6, Czech Republic}
\newcommand{\dapnia}{Dapnia, CEA Saclay, F-91191, Gif-sur-Yvette, France}
\newcommand{\debrecen}{Debrecen University, H-4010 Debrecen, Egyetem t{\'e}r 1, Hungary}
\newcommand{\elte}{ELTE, E{\"o}tv{\"o}s Lor{\'a}nd University, H - 1117 Budapest, P{\'a}zm{\'a}ny P. s. 1/A, Hungary}
\newcommand{\ewha}{Ewha Womans University, Seoul 120-750, Korea}
\newcommand{\fit}{Florida Institute of Technology, Melbourne, Florida 32901, USA}
\newcommand{\fsu}{Florida State University, Tallahassee, Florida 32306, USA}
\newcommand{\gsu}{Georgia State University, Atlanta, Georgia 30303, USA}
\newcommand{\hiroshima}{Hiroshima University, Kagamiyama, Higashi-Hiroshima 739-8526, Japan}
\newcommand{\ihepprot}{IHEP Protvino, State Research Center of Russian Federation, Institute for High Energy Physics, Protvino, 142281, Russia}
\newcommand{\illuiuc}{University of Illinois at Urbana-Champaign, Urbana, Illinois 61801, USA}
\newcommand{\inrras}{Institute for Nuclear Research of the Russian Academy of Sciences, prospekt 60-letiya Oktyabrya 7a, Moscow 117312, Russia}
\newcommand{\instpasczech}{Institute of Physics, Academy of Sciences of the Czech Republic, Na Slovance 2, 182 21 Prague 8, Czech Republic}
\newcommand{\isu}{Iowa State University, Ames, Iowa 50011, USA}
\newcommand{\jaea}{Advanced Science Research Center, Japan Atomic Energy Agency, 2-4 Shirakata Shirane, Tokai-mura, Naka-gun, Ibaraki-ken 319-1195, Japan}
\newcommand{\jinrdubna}{Joint Institute for Nuclear Research, 141980 Dubna, Moscow Region, Russia}
\newcommand{\jyvaskyla}{Helsinki Institute of Physics and University of Jyv{\"a}skyl{\"a}, P.O.Box 35, FI-40014 Jyv{\"a}skyl{\"a}, Finland}
\newcommand{\kek}{KEK, High Energy Accelerator Research Organization, Tsukuba, Ibaraki 305-0801, Japan}
\newcommand{\korea}{Korea University, Seoul, 136-701, Korea}
\newcommand{\kurchatov}{Russian Research Center ``Kurchatov Institute", Moscow, 123098 Russia}
\newcommand{\kyoto}{Kyoto University, Kyoto 606-8502, Japan}
\newcommand{\labllr}{Laboratoire Leprince-Ringuet, Ecole Polytechnique, CNRS-IN2P3, Route de Saclay, F-91128, Palaiseau, France}
\newcommand{\lahorelums}{Physics Department, Lahore University of Management Sciences, Lahore, Pakistan}
\newcommand{\lawllnl}{Lawrence Livermore National Laboratory, Livermore, California 94550, USA}
\newcommand{\losalamos}{Los Alamos National Laboratory, Los Alamos, New Mexico 87545, USA}
\newcommand{\lpc}{LPC, Universit{\'e} Blaise Pascal, CNRS-IN2P3, Clermont-Fd, 63177 Aubiere Cedex, France}
\newcommand{\lund}{Department of Physics, Lund University, Box 118, SE-221 00 Lund, Sweden}
\newcommand{\maryland}{University of Maryland, College Park, Maryland 20742, USA}
\newcommand{\mass}{Department of Physics, University of Massachusetts, Amherst, Massachusetts 01003-9337, USA }
\newcommand{\michigan}{Department of Physics, University of Michigan, Ann Arbor, Michigan 48109-1040, USA}
\newcommand{\muenster}{Institut fur Kernphysik, University of Muenster, D-48149 Muenster, Germany}
\newcommand{\muhlenberg}{Muhlenberg College, Allentown, Pennsylvania 18104-5586, USA}
\newcommand{\myongji}{Myongji University, Yongin, Kyonggido 449-728, Korea}
\newcommand{\nagasaki}{Nagasaki Institute of Applied Science, Nagasaki-shi, Nagasaki 851-0193, Japan}
\newcommand{\newmex}{University of New Mexico, Albuquerque, New Mexico 87131, USA }
\newcommand{\nmsu}{New Mexico State University, Las Cruces, New Mexico 88003, USA}
\newcommand{\ohio}{Department of Physics and Astronomy, Ohio University, Athens, Ohio 45701, USA}
\newcommand{\ornl}{Oak Ridge National Laboratory, Oak Ridge, Tennessee 37831, USA}
\newcommand{\orsay}{IPN-Orsay, Universite Paris Sud, CNRS-IN2P3, BP1, F-91406, Orsay, France}
\newcommand{\peking}{Peking University, Beijing 100871, P.~R.~China}
\newcommand{\pnpi}{PNPI, Petersburg Nuclear Physics Institute, Gatchina, Leningrad region, 188300, Russia}
\newcommand{\riken}{RIKEN Nishina Center for Accelerator-Based Science, Wako, Saitama 351-0198, Japan}
\newcommand{\rikjrbrc}{RIKEN BNL Research Center, Brookhaven National Laboratory, Upton, New York 11973-5000, USA}
\newcommand{\rikkyo}{Physics Department, Rikkyo University, 3-34-1 Nishi-Ikebukuro, Toshima, Tokyo 171-8501, Japan}
\newcommand{\saispbstu}{Saint Petersburg State Polytechnic University, St. Petersburg, 195251 Russia}
\newcommand{\saopaulo}{Universidade de S{\~a}o Paulo, Instituto de F\'{\i}sica, Caixa Postal 66318, S{\~a}o Paulo CEP05315-970, Brazil}
\newcommand{\seoulnat}{Seoul National University, Seoul, Korea}
\newcommand{\stonybrkc}{Chemistry Department, Stony Brook University, SUNY, Stony Brook, New York 11794-3400, USA}
\newcommand{\stonycrkp}{Department of Physics and Astronomy, Stony Brook University, SUNY, Stony Brook, New York 11794-3400, USA}
\newcommand{\subatech}{SUBATECH (Ecole des Mines de Nantes, CNRS-IN2P3, Universit{\'e} de Nantes) BP 20722 - 44307, Nantes, France}
\newcommand{\tenn}{University of Tennessee, Knoxville, Tennessee 37996, USA}
\newcommand{\titech}{Department of Physics, Tokyo Institute of Technology, Oh-okayama, Meguro, Tokyo 152-8551, Japan}
\newcommand{\tsukuba}{Institute of Physics, University of Tsukuba, Tsukuba, Ibaraki 305, Japan}
\newcommand{\vandy}{Vanderbilt University, Nashville, Tennessee 37235, USA}
\newcommand{\waseda}{Waseda University, Advanced Research Institute for Science and Engineering, 17 Kikui-cho, Shinjuku-ku, Tokyo 162-0044, Japan}
\newcommand{\weizmann}{Weizmann Institute, Rehovot 76100, Israel}
\newcommand{\wigner}{Institute for Particle and Nuclear Physics, Wigner Research Centre for Physics, Hungarian Academy of Sciences (Wigner RCP, RMKI) H-1525 Budapest 114, POBox 49, Budapest, Hungary}
\newcommand{\yonsei}{Yonsei University, IPAP, Seoul 120-749, Korea}
\affiliation{\abilene}
\affiliation{\acadsin}
\affiliation{\augie}
\affiliation{\banaras}
\affiliation{\barc}
\affiliation{\baruch}
\affiliation{\bnlcoll}
\affiliation{\bnlphys}
\affiliation{\caucr}
\affiliation{\charlesczech}
\affiliation{\chonbuk}
\affiliation{\ciae}
\affiliation{\cns}
\affiliation{\colorado}
\affiliation{\columbia}
\affiliation{\czechtech}
\affiliation{\dapnia}
\affiliation{\debrecen}
\affiliation{\elte}
\affiliation{\ewha}
\affiliation{\fit}
\affiliation{\fsu}
\affiliation{\gsu}
\affiliation{\hiroshima}
\affiliation{\ihepprot}
\affiliation{\illuiuc}
\affiliation{\inrras}
\affiliation{\instpasczech}
\affiliation{\isu}
\affiliation{\jaea}
\affiliation{\jinrdubna}
\affiliation{\jyvaskyla}
\affiliation{\kek}
\affiliation{\korea}
\affiliation{\kurchatov}
\affiliation{\kyoto}
\affiliation{\labllr}
\affiliation{\lahorelums}
\affiliation{\lawllnl}
\affiliation{\losalamos}
\affiliation{\lpc}
\affiliation{\lund}
\affiliation{\maryland}
\affiliation{\mass}
\affiliation{\michigan}
\affiliation{\muenster}
\affiliation{\muhlenberg}
\affiliation{\myongji}
\affiliation{\nagasaki}
\affiliation{\newmex}
\affiliation{\nmsu}
\affiliation{\ohio}
\affiliation{\ornl}
\affiliation{\orsay}
\affiliation{\peking}
\affiliation{\pnpi}
\affiliation{\riken}
\affiliation{\rikjrbrc}
\affiliation{\rikkyo}
\affiliation{\saispbstu}
\affiliation{\saopaulo}
\affiliation{\seoulnat}
\affiliation{\stonybrkc}
\affiliation{\stonycrkp}
\affiliation{\subatech}
\affiliation{\tenn}
\affiliation{\titech}
\affiliation{\tsukuba}
\affiliation{\vandy}
\affiliation{\waseda}
\affiliation{\weizmann}
\affiliation{\wigner}
\affiliation{\yonsei}
\author{A.~Adare} \affiliation{\colorado}
\author{S.~Afanasiev} \affiliation{\jinrdubna}
\author{C.~Aidala} \affiliation{\mass} \affiliation{\michigan}
\author{N.N.~Ajitanand} \affiliation{\stonybrkc}
\author{Y.~Akiba} \affiliation{\riken} \affiliation{\rikjrbrc}
\author{H.~Al-Bataineh} \affiliation{\nmsu}
\author{J.~Alexander} \affiliation{\stonybrkc}
\author{A.~Angerami} \affiliation{\columbia}
\author{K.~Aoki} \affiliation{\kyoto} \affiliation{\riken}
\author{N.~Apadula} \affiliation{\stonycrkp}
% ...
\collaboration{PHENIX Collaboration} \noaffiliation

\date{\today}


\begin{abstract}

Measurements of transverse-single-spin asymmetries ($A_{N}$) in $p$$+$$p$ 
collisions at $\sqrt{s}=62.4$ and 200~GeV with the PHENIX detector at RHIC 
are presented.  At midrapidity, $A_{N}$ is measured for neutral pion and eta 
mesons reconstructed from diphoton decay, and, at forward rapidities, 
neutral pions are measured using both diphotons and electromagnetic 
clusters.  The neutral-pion measurement of $A_{N}$ at midrapidity is 
consistent with zero with uncertainties a factor of 20 smaller than 
previous publications, which will lead to improved constraints on the 
gluon Sivers function.  At higher rapidities, where the valence quark 
distributions are probed, the data exhibit sizable asymmetries.  In 
comparison with previous measurements in this kinematic region, the new 
data extend the kinematic coverage in $\sqrt{s}$ and $p_T$, and it is 
found that the asymmetries depend only weakly on $\sqrt{s}$. The origin of 
the forward $A_{N}$ is presently not understood quantitatively.  The 
extended reach to higher $p_T$ probes the transition between transverse 
momentum dependent effects at low $p_T$ and multi-parton dynamics at high 
$p_T$.

\end{abstract}

\maketitle


\section{Introduction}
\label{sec:introduction}
The proton is a fundamental and stable bound state of quantum 
chromodynamics.  Collinear perturbative quantum chromodynamics (pQCD) at 
leading twist in the operator product expansion successfully describes the 
quark and gluon substructure of the proton observed in high energy 
scattering experiments~\cite{CTEQ10}.  The parton distribution functions, 
$f_{i}(x,Q^{2})$, constitute the number densities of partons of flavor $i$ 
in the proton.  They depend on the partonic momentum fraction, $x$, and on 
the momentum transfer scale, $Q^{2}$.  Two similar sets of distribution 
functions parametrize the spin dependent parton distributions in protons 
polarized either longitudinally or transversely with respect to the proton 
momentum direction~\cite{Ralston:1979tr}. The longitudinally polarized 
structure has been successfully described using pQCD at leading 
twist~\cite{DSSV}.

Initially, transverse-single-spin asymmetries or the analyzing power 
($A_{N}$) of hadrons $h$ produced in the transversely polarized 
$p^{\uparrow}+p \rightarrow h+X$ reaction were expected to be 
small~\cite{PhysRevLett.41.1689}, but experiments instead measured large 
asymmetries of up to $A_{N} \approx 40\%$. These asymmetries have been 
measured at increasing center-of-mass energies $\sqrt{s}$ over the past 
three decades, from 4.9 to 
200~GeV~\cite{Klem:1976ui,Antille:1980th,Adams:1991rw,Adams:1991cs,Allgower:2002qi,Arsene:2008mi}, 
Recent results from the Relativistic Heavy Ion Collider (RHIC) show that 
large asymmetries persist even up to 
$\sqrt{s}=200$~GeV~\cite{Abelev:2008,Adamczyk:2012qj,PhysRevD.86.051101}. 
Unlike at the low to intermediate energies, the measured unpolarized cross 
section at high energies is well reproduced by pQCD 
calculations~\cite{Abelev:2008, PhysRevD.86.051101}, indicating that 
unpolarized collisions can be described by the standard collinear 
factorized theory, while transversely polarized collisions cannot.

To better describe the large $A_{N}$ measurements, the theoretical 
framework has been extended to include transverse momentum dependent (TMD) 
distributions and multi-parton dynamics (higher twist effects). Because 
the intrinsic partonic transverse momentum scale is set by the mass of the 
proton, these effects dominate for hadrons with low momenta transverse to 
the beam axis, $p_{T}\lesssim 1$~GeV/$c$. At least two TMD effects have 
been proposed to explain the observed nonzero asymmetries.

The first of these, known as the Sivers effect, correlates the proton spin 
with the partonic transverse momentum $k_{T}$~\cite{Sivers:1989cc}. It has 
been measured in semi-inclusive deep inelastic scattering (SIDIS) 
experiments with sensitivity mainly to the 
quarks~\cite{Adolph:2012383,Airapetian:2009ti}. Previous results in 
$p$+$p$ collisions~\cite{Adler:2005in} have been used to constrain the 
gluon Sivers function~\cite{Anselmino:2006yq}.  Recently, this function 
has received intense theoretical attention based on questions of 
universality and an expected sign change of $A_{N}$ in SIDIS compared to 
Drell-Yan production~\cite{Collins:2002,Kang:2011hk}.

A second transverse momentum dependent effect, known as the Collins 
effect, describes the coupling of a transverse quark polarization 
(transversity) and a transverse spin dependent fragmentation from a struck 
quark into a hadron~\cite{Collins:1992kk}.  The spin dependent 
fragmentation part has been measured in $e^{+}$+$e^{-}$ annihilation for 
charged pions~\cite{Abe:2005zx,Seidl:2008xc} and serves as input for 
Collins asymmetries in proton scattering to access the transversity 
distribution~\cite{Airapetian:2010ds,Adolph:2012376,Anselmino:2007fs}. The 
full integral over all partonic momenta $0\leq x\leq 1$ of the 
transversity distribution can be compared to the flavor-singlet tensor 
charge $\delta\Sigma$, which is calculable in lattice 
QCD~\cite{PhysRevD.56.433,PhysRevD.79.014033}. This will be a fundamental 
test of the theory.

At large transverse momenta the collinear higher twist effects are thought 
to become more important in the creation of transverse spin 
asymmetries~\cite{Efremov:1984ip,Qiu:1991pp}. For Drell-Yan production it 
has been shown that both initial state TMD and multi-parton dynamics 
provide equivalent descriptions of transverse asymmetries in an overlap 
region at intermediate transverse momenta~\cite{Ji:2006ub}. With 
increasing $p_{T}$ the asymmetries are expected to fall off and vanish in 
the strictly collinear regime, which has not been observed experimentally 
yet. It is best probed at high center-of-mass energies, where the range of 
transverse momenta is wider.

This paper reports on measurements of $A_{N}$ at $\sqrt{s}=62.4$ and 
200~GeV. The data were taken by the PHENIX experiment at RHIC in the years 
2006 ($\sqrt{s}=62.4$~GeV) and 2008 ($\sqrt{s}=200$~GeV) with integrated 
luminosities of 42~nb$^{-1}$ and 4.3~pb$^{-1}$, respectively. Results are 
presented for neutral mesons in a midrapidity region ($|\eta|<0.35$) as 
well as for $\pi^{0}$ mesons and inclusive electromagnetic clusters at 
forward/backward pseudorapidities ($3.1<|\eta|<3.8$). 
Section~\ref{sec:detector} describes the experimental setup along with the 
properties of the polarized proton beams. The formalism of transverse 
single spin asymmetries is introduced in Sec.~\ref{sec:analysis} before 
details of the analysis procedure are specified. A general discussion of 
the results and their possible implications for nucleon structure and 
contributing asymmetry mechanisms concludes this paper in 
Sec.~\ref{sec:discussion}.



\section{Experimental Setup}
\label{sec:detector}

\subsection{PHENIX Midrapidity and Global Detectors}

The PHENIX midrapidity spectrometer is used to detect neutral pions and 
$\eta$ mesons via their decay into two photons. The spectrometer covers a 
pseudorapidity range of $|\eta|<0.35$ and is split into two approximately 
back-to-back arms each covering $\Delta \varphi=\pi/2$ in azimuth. The 
electromagnetic calorimeter (EMCal) is highly segmented with $\Delta \eta 
\times \Delta \varphi \approx 0.01 \times 0.01$. Events are selected using 
an EMCal based high tower energy trigger in coincidence with a minimum 
bias trigger. The trigger, digitization electronics, and details of the 
hardware have been discussed previously~\cite{Aphecetche:2003zr}. The 
trigger efficiency starts at about 5\% for neutral pions with 
$p_{T}\approx 1$~GeV/$c$ and rises to and saturates at about 90\% at 
$p_{T}>3.5$~GeV/$c$. A multi-wire proportional chamber with pad 
readout~\cite{Adcox:2003zp} is situated in front of the calorimeter face, 
and is used to veto charged particles.

The minimum bias trigger was defined as the coincidence of signals from 
two Beam Beam Counters (BBC) covering the full azimuthal angle and the 
pseudorapidity range $3.0<|\eta|<3.9$~\cite{Allen:2003zt}.  The BBCs are 
used to reconstruct the collision time and the collision (vertex) position 
along the beam direction.  Each BBC is situated 144~cm from the nominal 
interaction point and comprises an array of 64 counters arranged around 
the beam pipe.  Each counter comprises a \v{C}erenkov quartz-radiator of 
hexagonal cross section with a mesh dynode photomultiplier tube for 
read-out.

\subsection{Muon Piston Calorimeter}

The PHENIX Muon Piston Calorimeter (MPC) is an electromagnetic calorimeter 
which was designed to measure photons and neutral mesons at 
forward-rapidity. The detector comprises two separate devices placed along 
the beamline to the North and to the South of the nominal interaction 
point, labeled N-MPC and S-MPC. The S-MPC was first installed in 2006 and 
the N-MPC followed a year later. Therefore, the analysis of the 2006 data 
set ($\sqrt{s}$=62.4~GeV) uses only the S-MPC while the 2008 data set 
($\sqrt{s}$=200~GeV) includes both detectors.

\begin{figure}[bth]
  \centering
  \includegraphics[width=1.0\linewidth]{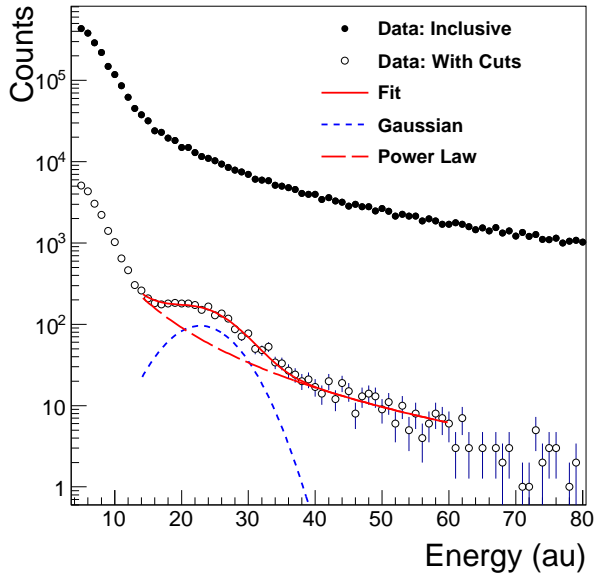}
  \caption{(color online)
Uncalibrated energy spectra with and without cuts to isolate minimum 
ionizing particles (MIP) in the MPC.  These cuts include: neighboring 
tower energy deposits and track-matching cuts using the upstream BBCs 
\v{C}erenkov counters.  The spectrum with the cut is fit with a power law 
and a Gaussian. The Gaussian peak position is taken as the most probable 
MIP energy deposit, $E\approx 234$~MeV.
}
\label{fig:mpc_mip_fit}
\end{figure}

Both MPCs are located in cavities of the steel piston which is part of the 
PHENIX muon detector magnet yoke. The diameter of each cavity limits the 
detector's outer diameter to 45~cm, while the beampipe requires an inner 
diameter of no less than 8~cm (N-MPC) or 10~cm (S-MPC).  The MPCs are 
placed $\pm$220~cm from the nominal interaction point and are composed of 
192 (S-MPC) or 220 (N-MPC) towers stacked to form an annulus around the 
beampipe. The detector acceptance covers the full azimuthal angle and a 
pseudorapidity range of $-3.8<\eta<-3.1$ South and $3.1<\eta<3.9$ North of 
the nominal interaction point.

Each tower combines a PbWO$_{4}$ scintillating crystal wrapped with 
Tyvek$\textsuperscript{\textregistered}$, aluminized mylar and 
MonoKote$\textsuperscript{\textregistered}$, with a Hamamatsu S8664-55 
avalanche photodiode for read-out. Each crystal measures 
2.2$\times$2.2$\times$18~cm$^{3}$, corresponding to a depth of 21.2 
radiation lengths and 0.844 nuclear interaction lengths.  Independently of 
the minimum bias trigger, the MPC is equipped with its own high energy 
cluster trigger. The trigger and digitization electronics are identical to 
those of the EMCal and are discussed in detail 
in~\cite{Aphecetche:2003zr}. For the presented data, the trigger 
efficiency starts at 5\% for photon energies $E\approx 30$~GeV and reaches 
a plateau at 90\% above $E>50$~GeV.

\begin{figure}[bth]
  \includegraphics[width=0.99\linewidth]{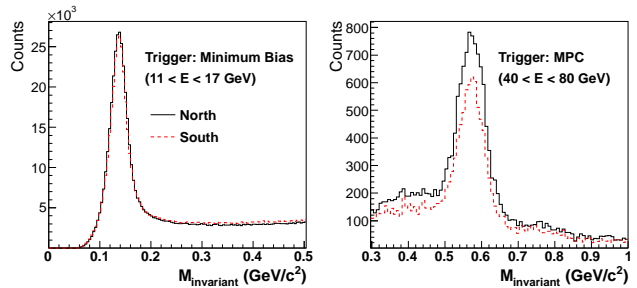}
  \caption{(color online)
Two-cluster invariant mass distributions from the 2008 data set at 
$\sqrt{s}$=200~GeV for both the North and South MPC detectors. The left 
panel shows $\pi^{0}$ peak from minimum bias triggered data set at low 
energy while the right side shows the $\eta$ meson peak from the MPC 
triggered data set at high two-cluster energies $E$. A comparison of the 
peak position and widths from data and simulation are used to determine 
the energy scale uncertainty. }
  \label{fig:mpc_2gamma}
\end{figure}

A test-beam measurement, carried out at the Meson Test Beam Facility 
\footnote{now the MT6 area at the Fermilab Test Beam Facility} at the 
Fermi National Accelerator Laboratory confirmed the calorimeter's linear 
energy response and measured the electromagnetic shower shapes. These 
shower shapes were then used to tune a {\sc geant} 3.21~\cite{Geant:1987} 
based full detector simulation. The absolute energy scale of the detector 
is determined \emph{in situ} using a two-step process.  First, minimum 
ionizing particles are used. Yields of charged tracks in the calorimeter 
are enhanced by requiring a correlated hit in the BBC that is located in 
front of the MPC. Additionally, the tower multiplicity of the cluster is 
required to be small compared to a typical electromagnetic shower to 
increase the hadronic contributions. A sample minimum ionizing particle 
peak is shown in Fig.~\ref{fig:mpc_mip_fit} with an expected mean energy 
of 234~MeV. The initial MIP calibration is then used as the seed in an 
iterative and converging procedure for individual towers that is based on 
the $\pi^{0}$ peak in the invariant mass distribution. Time dependencies 
in the tower gains are tracked and corrected for by a monitoring system of 
LEDs, whose intensities are monitored by PIN diodes. Finally, the overall 
calibration is verified and the energy resolution is determined by 
comparing the masses of the $\pi^{0}$ and $\eta$ peaks in the two-cluster 
invariant mass distributions between data and a Monte-Carlo simulation. A 
set of representative two-cluster invariant mass peaks is shown in 
Fig.~\ref{fig:mpc_2gamma}. The relative energy resolution ($\delta E/E$) 
of the calorimeter is found to be 13\%/$\sqrt{E}\oplus$8\% with an overall 
energy scale uncertainty of 2\%.

\subsection{Polarized Proton Beams}
\label{sec:polarized_beams}

RHIC accelerates and stores polarized proton beams at energies up to 
255~GeV in two independent rings.  The beams collide at several 
interaction points along the ring.  Each ring can be filled with up to 120 
bunches with different transverse polarization directions.  These 
directions alternate to reduce systematic effects from slow variations in 
luminosity or detector acceptances and efficiencies. Additionally, the 
patterns are chosen from four predefined basic patterns to reduce time 
dependent correlations and detector effects.

Previous publications describe in detail the necessary accelerator 
instrumentation for producing the colliding polarized 
beams~\cite{Alekseev:2003sk}. The polarization is measured with a set of 
polarimeters external to the PHENIX experiment using elastic scattering 
from a hydrogen gas jet or a Carbon fiber target. For the determination of 
the absolute polarization of both proton beams, the hydrogen jet 
polarimeter with a known polarization of the atomic jet is 
used~\cite{PhysRevD.79.094014}. Due to the low density of the gas jet a 
polarization measurement with good accuracy requires many hours of data 
taking. Therefore, the relative polarization is measured several times per 
fill with high precision by fast $p$+C polarimeters for each of the two 
storage rings~\cite{Nakagawa.AIP2008}. These relative measurements are 
then normalized using results from the jet polarimeter.

While both of the RHIC beams are polarized during the measurement of the 
single spin asymmetries presented in this paper, summation over the 
bunches of one beam effectively averages the polarization to zero.  This 
procedure is applied to one of the two beams at a time and can therefore 
be used as a cross check of two uncorrelated measurements of the 
asymmetry. The direction of the polarized beam is commonly referred to as 
forward in the following; backward is in the direction of the unpolarized 
beam. Table~\ref{tbl:beampol} summarizes the beam polarizations for the 
different data sets and center-of-mass energies, with $\vec{p}=(0,0,p_z)$ 
pointing North, according to the PHENIX coordinate system.

\begin{table}[ht]
  \caption{Polarizations for RHIC proton beams in 2006 and 2008.
The polarization uncertainty is a global scale uncertainty of the measured 
asymmetries $A_{N}$ and is not included in any of the figures or data 
tables.
  }
  \label{tbl:beampol}

  \begin{ruledtabular}\begin{tabular}{cccc}
      Year & $\sqrt{s}$ (GeV) & Beam direction & $P_{beam}$ \\
      \hline
      2006 & 62.4 & North, $\vec{p}=(0,0,p_z)$  & (49.0$\pm$4.4)\% \\
      2006 & 62.4 & South, $\vec{p}=(0,0,-p_z)$ & (49.0$\pm$4.4)\% \\
      2008 & 200  & North, $\vec{p}=(0,0,p_z)$  & (48.0$\pm$4.0)\% \\
      2008 & 200  & South, $\vec{p}=(0,0,-p_z)$ & (41.0$\pm$4.0)\% \\
  \end{tabular}\end{ruledtabular}
\end{table}

The stable polarization direction around the accelerator is vertical 
($P^{\uparrow}=(0,P,0)$ or $P^{\downarrow}=(0,-P,0)$) and can be changed 
by spin rotators around the collision points. The transverse components of 
the polarization vector are measured locally in PHENIX. In 2008, the 
polarization of the North pointing beam was tilted from the vertical 
direction by $\varphi_{0}=0.263\pm0.030^{\rm stat}\pm0.090^{\rm syst}$ 
rad. For the rest of the measurements, all other polarization vectors are 
found to be consistent with the vertical direction within statistical 
uncertainties~\cite{Adare:2010bd}. The polarization directions are 
accounted for in the determination of the relevant asymmetries.  In 
addition, the 2006 and 2008 polarization direction measurements have been 
independently verified using the analysis techniques from 
Sec.~\ref{sec:formalism}.

\section{ANALYSIS}\label{sec:analysis}


\subsection{Transverse-Single-Spin Asymmetries}
\label{sec:formalism}

The $A_{N}$ that can generally arise in polarized scattering experiments are 
described in the framework of polarization analyzing tensors which gives 
information about fully polarized initial and final states of the 
scattering process. The polarization can be aligned along three dimensions 
in the scattering frame, i.e., longitudinal in the projectile direction 
$\vec{L}$, sideways in the scattering (or production) plane $\vec{S}$, or 
normal to the scattering plane $\vec{N}$, where 
$\vec{S}=\vec{N}\times\vec{L}$. In the following, the left side refers to 
the direction of $\vec{S}$ in this right-handed system, the right side to 
the opposite direction.  For $A_{N}$, we are only considering a normal 
polarization for the projectile. Target and final states are unpolarized. 
The normal space quantization can create a transverse asymmetry within the 
scattering plane. A rotation into the laboratory frame (where the beam 
polarization $P$ is prepared) then transforms this pure left-right 
asymmetry into an azimuthal ($\varphi$) modulation of the cross section 
$d\sigma(\varphi)\propto A_{N}\cdot P\cdot\cos\varphi$. The transverse 
asymmetry $A_{N}$ can be determined from point-like detectors as:
\begin{equation}
  A_{N} = \frac{1}{P} \cdot \frac{1}{\cos\varphi}
  \frac{d\sigma(\varphi)-d\sigma(\varphi+\pi)}{d\sigma(\varphi)+d\sigma(\varphi+\pi)}.
  \label{eq:asym2}
\end{equation}

The same result can be achieved with a detector in just one hemisphere 
by a rotation of the polarization vector 
$P^{\uparrow}\rightarrow P^{\downarrow}$:
\begin{eqnarray*}
  d\sigma^{\uparrow}(\varphi) & = & d\sigma^{\downarrow}(\varphi+\pi) \\
  d\sigma^{\downarrow}(\varphi) & = & d\sigma^{\uparrow}(\varphi+\pi).
\end{eqnarray*}

Integrating the cross sections over the detector acceptance,
beam luminosities, and the duration of the measurement, $A_{N}$ is
experimentally extracted from the geometric means of the particle 
yields:
\begin{equation}
  \epsilon(\varphi)= A_{N} \cdot P \cdot\cos\varphi =
  \frac{\sqrt{N^{\uparrow}_{L}\cdot N^{\downarrow}_{R}}-\sqrt{N^{\downarrow}_{L}\cdot N^{\uparrow}_{R}}}
       {\sqrt{N^{\uparrow}_{L}\cdot N^{\downarrow}_{R}}+\sqrt{N^{\downarrow}_{L}\cdot N^{\uparrow}_{R}}},
  \label{eq:asym}
\end{equation}
where $N_{L}$, $N_{R}$ refer to particle yields in detector segments 
$\Delta\varphi$ of the left ($\varphi$) and right ($\varphi+\pi$) 
hemispheres. An alternate estimator is used to study systematic effects
\begin{equation}
\epsilon(\varphi)=A_{N} \cdot P \cdot\cos\varphi =
\frac{N^{\uparrow}-\mathcal{R} \cdot N^{\downarrow}}{N^{\uparrow}+\mathcal{R} \cdot N^{\downarrow}},
\label{eq:lumi}
\end{equation}
with $\mathcal{R}$ being the ratio of luminosities between the two spin 
states $\uparrow$ and $\downarrow$.  This luminosity is determined using 
the polarization-sorted counts from the minimum bias trigger. The 
asymmetries in this analysis are calculated in 8 or 16 bins in the 
azimuth, unless noted otherwise, and then fit to the cosine modulation 
(with and without an additional free phase $\varphi_{0}$ for consistency 
checks).  Systematic uncertainties are estimated by comparing asymmetries 
from Eqs.~\ref{eq:asym} and \ref{eq:lumi}, which may be due to different 
assumptions in the integration of the cross sections.


\subsection{$A^{\pizero}_{N}$ at $\sqrt{s}=62.4$~GeV and High $x_{F}$}
\label{sec:mpc62}

Measurements at $\sqrt{s}=62.4$~GeV were carried out in 2006 with the 
South MPC, from a total of $3.6\times10^7$ MPC triggered events. The 
$\pi^0\rightarrow\gamma+\gamma$ decay is reconstructed from pairs of 
clusters in the detector with a selection on the photon shower shape.  
Clusters which have their central tower marked as either noisy or inactive 
are removed from the analysis. The $\pi^0$ contribution is selected from 
the cluster pairs by requiring a minimum pair energy, $E_{\rm 
pair}>6$~GeV, and an upper limit of 0.6 on the energy asymmetry $\alpha$ 
of the two cluster energies $E_{1}$ and $E_{2}$,
\begin{equation}
\alpha = \left|\frac{E_1 - E_2}{E_1+E_2}\right|.
\label{eqn:alpha}
\end{equation}

The two cluster invariant mass distributions look qualitatively similar to 
those from $\sqrt{s}=200$~GeV shown in Fig.~\ref{fig:mpc_2gamma}.  The 
shape of the distributions has been studied in simulations based on the 
{\sc pythia} event generator~\cite{PythiaManual:2006} Tune 
A~\cite{Skands:2010ak} with a full detector simulation (similar to 
Sec.~\ref{sec:detector}.B). The background is dominated by combinatorial 
effects from reconstructing two clusters from different parent sources. 
The background yield is determined by mixing uncorrelated clusters from 
different events and normalizing to the invariant mass distribution above 
the \pizero peak, but below any contribution from the $\eta$ peak.  From 
the integral of the resulting \pizero peak one can determine the $\pizero$ 
yields. The final asymmetries are calculated according to 
Eq.~\ref{eq:asym} from the geometrical means of the $\pizero$ yields.  
The systematic uncertainties to these asymmetries are estimated using 
Eq.~\ref{eq:lumi}.

\begin{figure}[htp]
  \centering
  \includegraphics[width=1.0\linewidth]{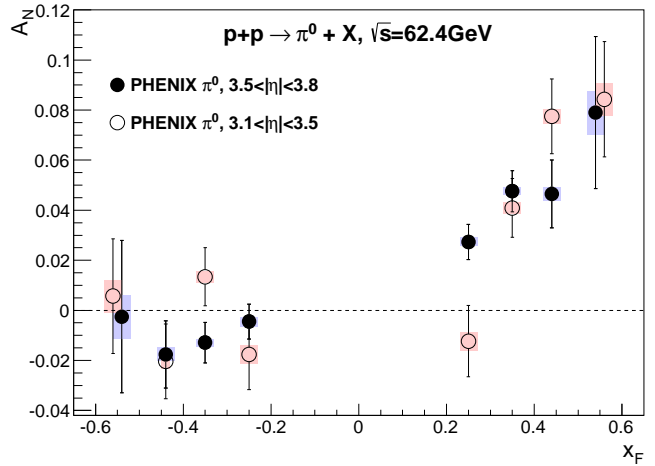}
  \caption{(color online)
Neutral pion $A_{N}$ at $\sqrt{s}=62.4$~GeV as function of $x_{F}$ in two 
different pseudorapidity ranges ($3.1<|\eta|<3.5$ and $3.5<|\eta|<3.8$) 
with statistical and systematic uncertainties.  Appendix 
Table~\ref{tab:mpc62:AN_xf_etacuts} gives the data in plain text. An 
additional uncertainty from the beam polarization (see 
Table~\ref{tbl:beampol}) is not included.
	}
  \label{fig:mpc62_xf}
\end{figure}

\begin{figure}[htp]
  \centering
  \includegraphics[width=1.0\linewidth]{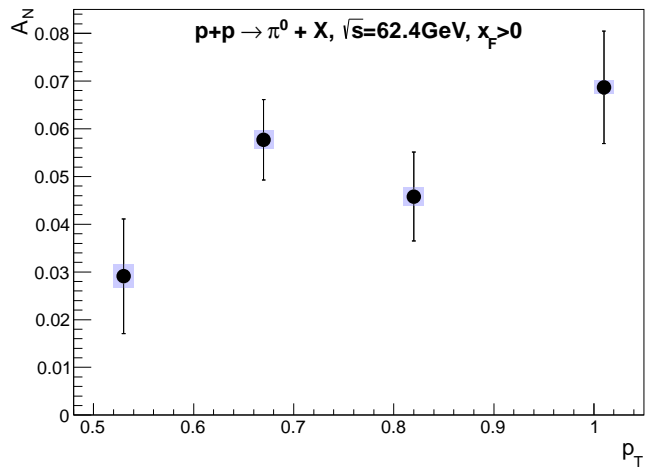}
  \caption{
Neutral pion $A_{N}$ at $\sqrt{s}=62.4$~GeV as function of transverse 
momentum $p_{T}$ Appendix Table~\ref{tab:mpc62:AN_pt} gives the plain text 
data. An additional uncertainty from the beam polarization (see 
Table~\protect\ref{tbl:beampol}) is not included.
	}
  \label{fig:mpc62_pt}
\end{figure}

Figure~\ref{fig:mpc62_xf} shows $A_{N}$ at $\sqrt{s}=62.4$~GeV as a function 
of $x_{F}=2\cdot p_{z}/\sqrt{s}$, with $p_{z}$ being the longitudinal 
component of the momentum along the direction of the polarized proton 
beam. While there is a significant, nonzero asymmetry rising with 
$x_{F}>0$ in the forward direction, no such behavior can be seen at 
negative $x_{F}<0$ where the asymmetries are consistent with zero.  
Figure~\ref{fig:mpc62_pt} shows the $p_{T}$ dependence of $A_{N}$, up to a 
range that is largely limited by kinematics due to the low 62.4 GeV 
center-of-mass energy.  No strong $p_{T}$ dependence is observed.

\begin{figure}[htp]
  \centering
  \includegraphics[width=1.0\linewidth]{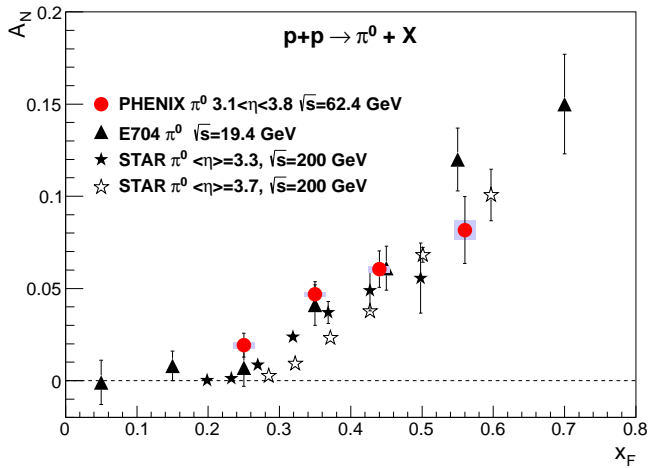}
  \caption{(color online)
Comparison of neutral pion $A_{N}$ as function of $x_{F}$ from 
$\sqrt{s}=19.4$ to 200~GeV from this publication 
and~\protect\cite{Adams:1991rw,Abelev:2008}. Appendix 
Table~\ref{tab:mpc62:AN_xf} gives the data in plain text.
	}
  \label{fig:mpc62_roots}
\end{figure}

Figure~\ref{fig:mpc62_roots} compares the $x_{F}$-dependence of neutral 
pion $A_{N}$ of this publication with the world data 
set~\cite{Abelev:2008,Arsene:2008mi} at center-of-mass energies from 
$\sqrt{s}=19.4$ to 200~GeV.  Although the different measurements were 
carried out with slightly different detector acceptances, there is a 
general agreement between the asymmetries in the onset of nonvanishing 
asymmetries and the $x_{F}$ dependence.  The asymmetries appear to be 
independent of the center-of-mass energy, including at high energies where 
the applicability of pQCD is well established at $\sqrt{s}=200$~GeV at 
$p_{T}>2$~ GeV/$c$.

\begin{figure}[htp]
  \centering
  \includegraphics[width=1.0\linewidth]{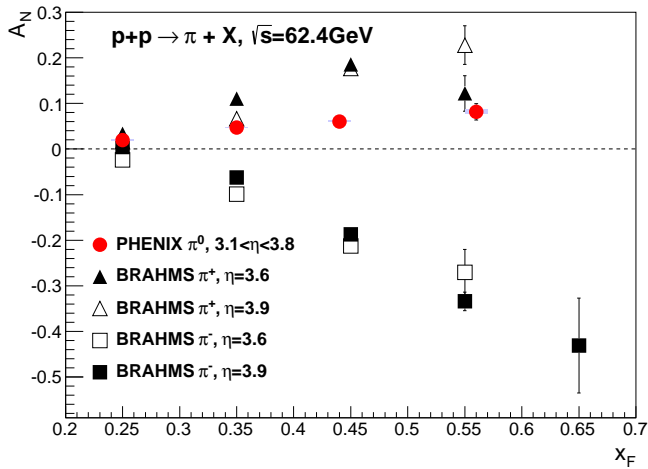}
  \caption{(color online)
Isospin comparison of pion $A_{N}$ as a function of $x_{F}$ at 
$\sqrt{s}=62.4$~GeV from this publication 
and~\protect\cite{Arsene:2008mi}. Appendix Table~\ref{tab:mpc62:AN_xf} 
gives the data in plain text.
}
\label{fig:mpc62_phenix_brahms}
\end{figure}

Figure~\ref{fig:mpc62_phenix_brahms} shows the pion isospin dependence of 
$A_{N}$ at $\sqrt{s}=62.4$~GeV with combined RHIC data from the new 
$\pi^{0}$ PHENIX data and charged pion data from the BRAHMS 
collaboration~\cite{Arsene:2008mi}.  The BRAHMS measurements of charged 
pions were carried out with two detector settings covering different 
subranges in pseudorapidity which compare well to the acceptance of the 
MPC.  While $\pi^{+}$ and $\pi^{0}$ asymmetries are positive, those of 
$\pi^{-}$ are of opposite sign.  The amplitudes of the charged pion 
asymmetries are of similar size, with the $\pi^{-}$ perhaps slightly 
larger, whereas both are significantly larger than the neutral pion 
asymmetry.

\subsection{$A^{cluster}_{N}$ at $\sqrt{s}=200$~GeV and High $x_{F}$}
\label{sec:mpc200}

At energies below $E_{\pi^0}\lesssim$ 20~GeV the MPC is able to resolve 
the $\pi^{0}\rightarrow\gamma+\gamma$ decay.  However, with increasing 
energy, the opening angle between the two photons becomes so small that 
their electromagnetic clusters fully merge in the detector. This limits 
the $x_{F}$ range at $\sqrt{s}=200$~GeV to below 0.2 for \pizero's 
reconstructed via the two-gamma decay mode.  To overcome this limitation 
the data analysis is done for inclusive clusters.

The data set at $\sqrt{s}=200$~GeV includes $1.8\times10^{8}$ events 
recorded with a high energy cluster trigger. Clusters in the analysis are 
required to have fired the corresponding trigger, i.e., N-MPC or S-MPC, 
and to satisfy a time of flight cut. Clusters whose central tower is 
either marked noisy or inactive are removed from the analysis. The 
contributions from hadrons to the cluster yields are reduced by selecting 
for photonic shower shapes. To minimize effects from energy leakage at the 
detector edges, a radial fiducial cut is applied. The transverse 
asymmetries are determined with Eq.~\ref{eq:asym} and systematic 
uncertainties are estimated using the difference from Eq.~\ref{eq:lumi}.

The cluster composition is estimated using Monte Carlo simulations.  
Again, a full detector simulation is based on input from {\sc pythia} 6.421 Tune A 
with separate normalization factors between direct photons ($k=2$) and all other 
particles originating from high energy scattering processes ($k=1$) with a minimum 
$p_T$ of 2~GeV.
The normalization factors are determined by comparing the simulated 
cross sections with RHIC measurements at $\sqrt{s}=200$~GeV 
\cite{xsect:star:pi0, xsect:brahms:h, xsect:phenix:dirgam, xsect:phnx:pi0}.
The composition analysis differentiates between electromagnetic
clusters originating from photonic decays of $\pi^0$ and $\eta$ mesons, 
direct photons, and energy deposited by charged hadrons ($h^{\pm}$).
Contributions from other sources, e.g.\ fragmentation photons and
$\omega$ meson decays, are combined in the ``other $\gamma$'' category.

\begin{figure}[ht]
  \includegraphics[width=1.0\linewidth]{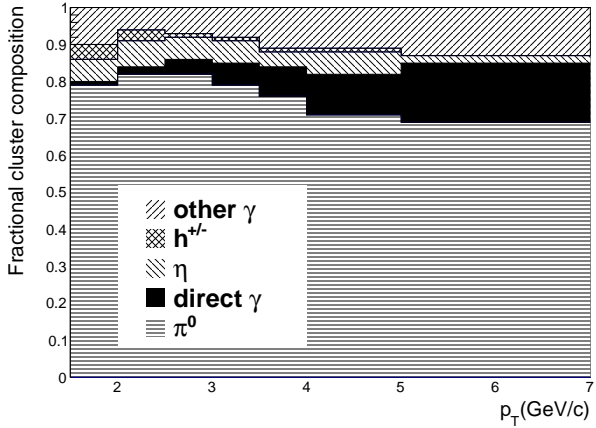}
  \caption{Cluster composition from $p$$+$$p$ Monte Carlo event generator studies
           at $\sqrt{s}=200$~GeV with a full detector simulation.
	   The kinematic cuts and $p_{T}$ ranges are the same as used in the data analysis and
	   directly comparable to Fig.~\ref{fig:mpc200:AN_pt}, in particular $x_{F}>0.4$.}
  \label{fig:ptDecomp}
\end{figure}

Figure \ref{fig:ptDecomp} summarizes the cluster composition as function of $p_{T}$ 
with large $x_{F}>0.4$; Table \ref{tbl:ClusterComposition} lists the corresponding
values in detail.
In the context of this {\sc pythia} study, over the studied kinematic range contributions
from decay photons of $\pi^{0}$ mesons 
are the dominant source of clusters in the MPC.
With increasing $p_{T}$ there is a sizable increase in contributions from direct and 
other photons.
The relative uncertainty of the composition from this study at $p_{T}>5$~GeV/$c$ is less than 20\% and 
significantly smaller at lower $p_{T}$.

\begin{figure}[tb]
  \includegraphics[width=1.0\linewidth]{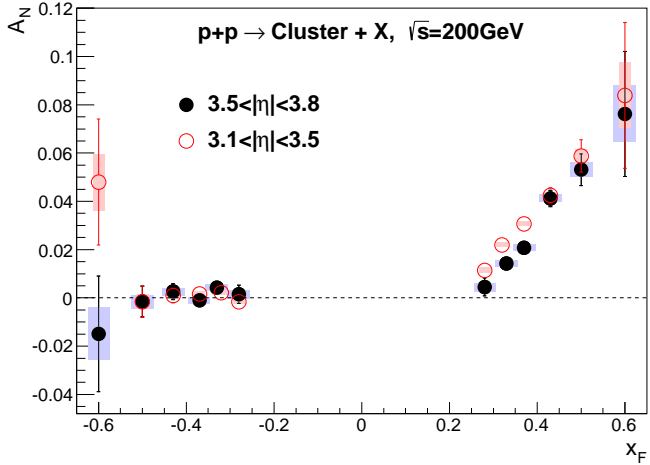}
  \caption{(color online)
The $A_{N}$ of electromagnetic clusters at $\sqrt{s}=200$~GeV as function 
of $x_{F}$ and in two different pseudorapidity ranges.  Appendix 
Table~\ref{tbl:mpc200:AN_xf} gives the data in plain text. An additional 
uncertainty from the beam polarization (see Table \ref{tbl:beampol}) is 
not included.
	   }
  \label{fig:mpc200:AN_xf}
\end{figure}

Figure~\ref{fig:mpc200:AN_xf} summarizes the $x_{F}$-dependence of the 
cluster $A_{N}$ for two different pseudorapidity ranges similar to 
Fig.~\ref{fig:mpc62_xf}. Systematic uncertainties again are evaluated by 
comparison of results from Eqs.~\ref{eq:asym} and \ref{eq:lumi}. Within 
statistical uncertainties the asymmetries in the backward direction 
$x_{F}<0$ are found to be consistent with zero, whereas in the forward 
direction $A_{N}$ rises almost linearly with $x_{F}$. The asymmetries are 
of similar size compared to earlier results at different center-of-mass 
energies as shown in Fig.~\ref{fig:mpc62_roots}.

\begin{figure}[thb]
  \includegraphics[width=1.0\linewidth]{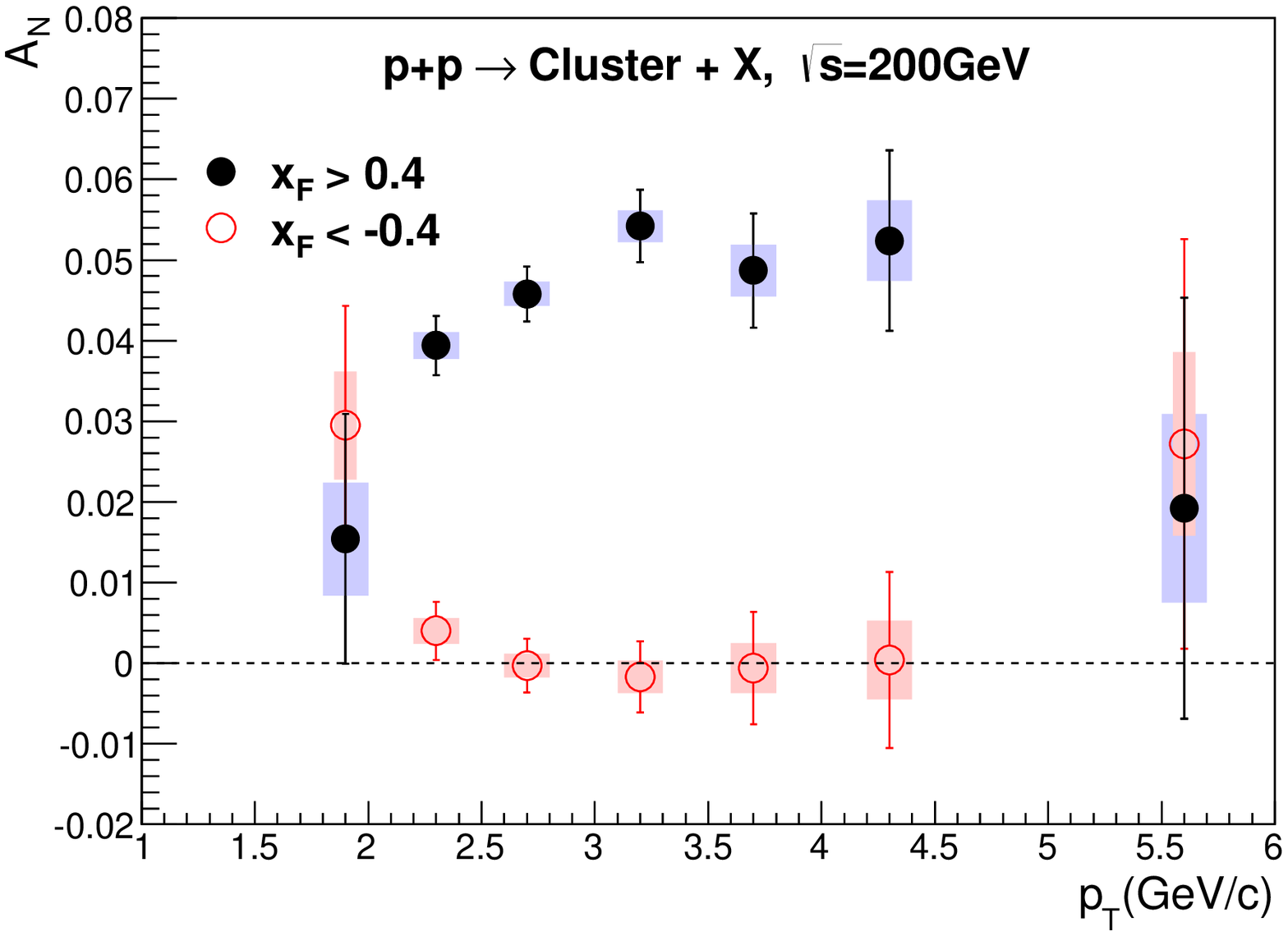}
  \caption{(color online)
The $A_{N}$ of electromagnetic clusters at $\sqrt{s}=200$~GeV at large 
$|x_{F}|>0.4$ in forward/backward directions as function of $p_{T}$. 
Appendix Table~\ref{tbl:mpc200:AN_pt} gives the data in plain text. An 
additional uncertainty from the beam polarization (see 
Table~\ref{tbl:beampol}) is not included.
	   }
  \label{fig:mpc200:AN_pt}
  \includegraphics[width=1.0\linewidth]{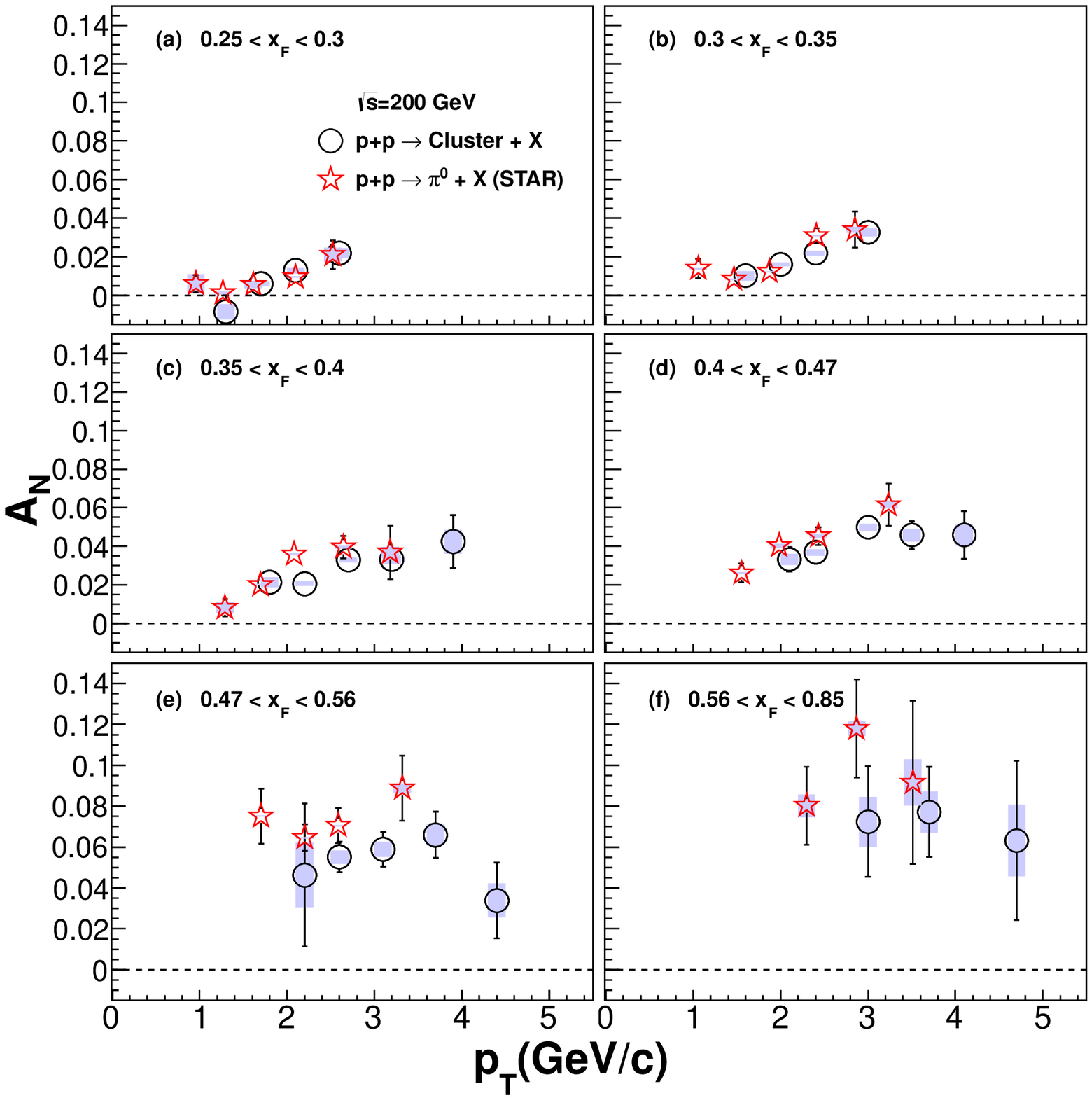}
    \caption{(color online)
Comparison of $A_{N}$ of electromagnetic clusters and $\pi^{0}$ 
mesons~\cite{Abelev:2008} at $\sqrt{s}=200$~GeV as function of $p_{T}$ in 
different ranges of $x_{F}$. Appendix Table~\ref{tbl:mpc200:AN_pt2} gives 
the data in plain text. An additional uncertainty from the beam 
polarization (see Table \ref{tbl:beampol}) is not included.
	     }
  \label{fig:mpc200:AN_pt2}
\end{figure}

Figure~\ref{fig:mpc200:AN_pt} presents $A_{N}$, as a function of transverse 
momentum $p_{T}$ for values of $|x_{F}|>0.4$ where $A_{N}$ is largest in 
forward kinematics (compare Fig.~\ref{fig:mpc200:AN_xf}). The asymmetry 
rises smoothly and then seems to saturate above $p_{T}>3$~GeV/$c$. A 
significant decrease of the asymmetry as expected from higher twist 
calculations is not observed~\cite{Ji:2006ub}. Again, negative $x_{F}$ 
asymmetries are found to be consistent with zero within statistical 
uncertainties.

Figure~\ref{fig:mpc200:AN_pt2} shows $A_{N}$ as a function of $p_{T}$ for 
different ranges of $x_{F}$.  These ranges are chosen to match that of an 
earlier measurement of $\pi^{0}$ asymmetries from the STAR 
experiment~\cite{Abelev:2008}.  The two measurements in general display a 
good agreement. At large $x_{F}$ and high $p_{T}$ there is perhaps a hint 
that the inclusive cluster asymmetries are smaller, but with present 
statistics the difference is not yet significant.  We note that the STAR 
measurement is for identified $\pi^{0}$'s and the PHENIX measurement is 
for clusters with a mixed composition.  As mentioned previously, these 
clusters are dominantly from $\pi^0$'s, but also include contributions 
from the decays of $\eta$ and other neutral mesons, as well as a 
contribution from direct photons which is increasing with $x_{F}$ and 
$p_{T}$.

\clearpage %

\subsection{$A^{\pi^{0},\eta}_{N}$ at $\sqrt{s}$=200 GeV and Small $x_{F}$}
\label{sec:emc200}

The data selection and asymmetry analysis in the midrapidity spectrometer
closely follows the procedure of previous analyses~\cite{Adler:2005in}.
The data set includes 6.9 $\times$ 10$^{8}$ events triggered by the
high $p_{T}$ photon trigger.  Photon clusters are selected using
photonic shower shape cuts in the electromagnetic calorimeter, the
time of flight between the collision point and the calorimeter, a
minimum deposited energy of 200~MeV, and a charged particle veto from
tracking in front of the calorimeter.  Cluster pairs are then chosen
with an energy asymmetry (Eq. \ref{eqn:alpha}) of less than 0.8 (0.7)
for $\pi^{0}$ ($\eta$) identification, and by requiring that the
photon with the higher energy fired the trigger.

The yields are taken as the number of cluster pairs in a $\pm$25
MeV/$c^{2}$ window around the mean of the $\pi^{0}$ peak in the
invariant mass distribution ($\pm$70~MeV/$c^{2}$ around the mean of
the $\eta$ mass).  The width of the $\pi^{0}$ peak decreases from 12
to 9~MeV/$c^{2}$ as $p_{T}$ increases from 1 to 12~GeV/$c$ (35 to 25~MeV/$c^{2}$ 
for the $\eta$).  The background fractions in the signal
windows depend on $p_{T}$ and range from 29\% to 4\% under the
$\pi^{0}$ peak and 75\% to 41\% for the $\eta$ peak as $p_{T}$ increases.

To remove a possible background asymmetry, the weighted
asymmetry between a low and high mass region around the signal peak is
determined and subtracted from the signal region.  These
regions are defined from 47 to 97 and from 177 to 227~MeV/$c^{2}$ for
the $\pi^{0}$, and from 300 to 400 and from 700 to 800~MeV/$c^{2}$ for
the $\eta$ meson.  The signal asymmetry $A_{N}^{\rm signal}$ can be
calculated using yields from the peak region $N_{\rm incl}$ and
from the interpolated background yields $N_{\rm bg}$:
\begin{equation}
A^{\rm signal}_{N}=\frac{A^{\rm incl}_{N}-r A^{\rm bg}_{N}}{1-r},
\label{eqn:back_subtract}
\end{equation}
with the background fraction $r=N_{\rm bg}/N_{\rm incl}$ under either the \
$\pi^{0}$ or $\eta$ signal.  The background asymmetries are all consistent with zero.

Due to the limited azimuthal acceptance of the midrapidity spectrometer
the asymmetries are only measured from integrated yields in the whole
detector hemispheres to the left and right of the polarization
direction.  To account for the cosine modulation of the particle
production, the asymmetries need to be corrected by an average factor
$f=1/\langle\cos\varphi\rangle$ taken over the detector acceptance.
The asymmetries are calculated from Eq.~\ref{eq:asym}, and the 
corresponding systematic uncertainties are estimated from differences 
with Eq.~\ref{eq:lumi}.

\begin{figure}[htp]
  \includegraphics[width=1.0\linewidth]{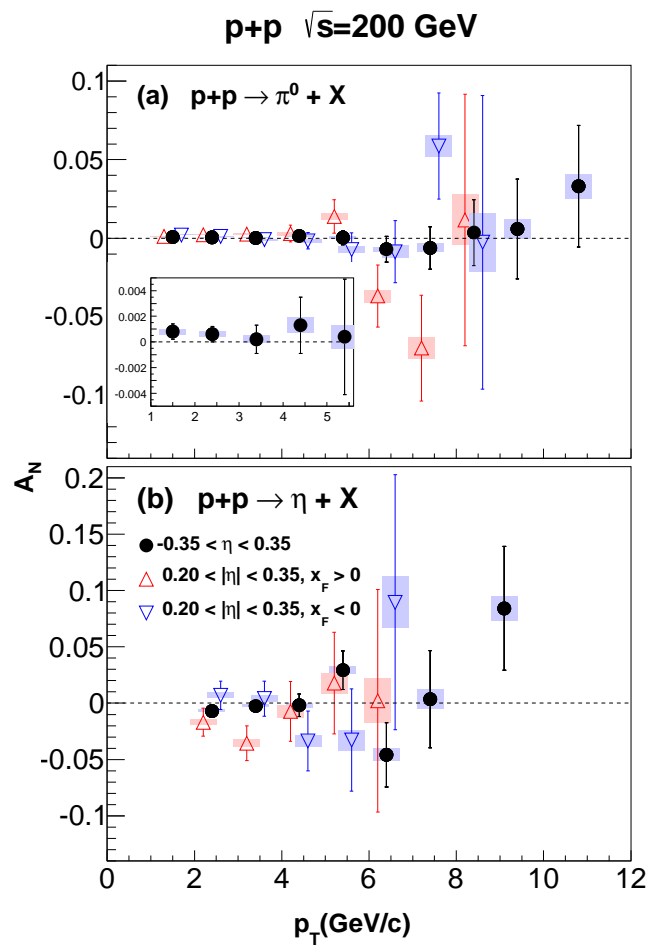}
  \caption{(color online)
The $A_{N}$ measured at midrapidity ($|\eta|<0.35$), as function of 
$p_{T}$ for $\pi^{0}$ (a) and $\eta$ (b) mesons (see 
Tables~\ref{tbl:emc200_pi0} and \ref{tbl:emc200_eta}). Triangles are 
slightly forward/backward going sub-samples of the full data set 
(circles). These are shifted in $p_{T}$ for better visibility. An 
additional uncertainty from the beam polarization (see 
Table~\ref{tbl:beampol}) is not included.
    }
  \label{fig:emc200}
\end{figure}

Both the inclusive and background asymmetries are determined for each 
RHIC fill to test for possible variations with time.
The mean values are then used for the calculation of the final asymmetries 
for $\pi^{0}$ and $\eta$ mesons as function of $p_{T}$, see Fig.~\ref{fig:emc200} 
and Tables \ref{tbl:emc200_pi0} and \ref{tbl:emc200_eta}.
The figure shows the asymmetries for the whole detector acceptance ($|\eta|<0.35$) 
and for two samples selecting slightly forward/backward going particles 
($0.2<|\eta|<0.35$).
It is important to note that the data in the restrictive pseudorapidity ranges
are sub-samples of the larger inclusive data set.
These very precise results are all consistent with zero over the observed $p_{T}$ 
range.


\section{Discussion} \label{sec:discussion}


The $A_{N}$ of neutral pions and inclusive charged hadrons have previously 
been measured with the PHENIX midrapidity spectrometer 
\cite{Adler:2005in}. Those asymmetries have been found to be consistent 
with zero and have been used to constrain the gluon Sivers 
function~\cite{Anselmino:2006yq} despite their limited statistical 
precision. The new results shown in Fig.~\ref{fig:emc200} exceed the 
former precision by a factor of 20 for the $\pi^{0}$ transverse 
asymmetries while extending the $p_{T}$ reach to above 10~GeV/$c$. Also, 
this paper reports on $A_{N}$ of $\eta$ mesons at $x_{F}\approx0$ which 
extends previous results~\cite{Adams1992531} both in $\sqrt{s}$ and 
$p_{T}$. Altogether, no significant deviation from zero can be seen in the 
results within the statistical uncertainties in the covered transverse 
momentum range. Any difference in the two meson asymmetries would likely 
be dominated by fragmentation effects. Either these are small or 
suppressed by the contributing transversity distribution in the covered 
kinematic range.

In the forward direction, nonvanishing meson asymmetries persist all the way up 
to $\sqrt{s}=200$~GeV, as shown in Figs.~\ref{fig:mpc62_xf} and 
\ref{fig:mpc200:AN_xf}.
While there is no asymmetry in the backward direction ($x_{F}<0$), $A_{N}$ scales 
almost linearly with positive $x_{F}>0.2$.
This behavior is similar to previous experimental results, as summarized 
in Fig.~\ref{fig:mpc62_roots}, where no strong center-of-mass energy dependence 
of the asymmetry is observed.
The kinematic coverage of the experiments is not exactly the same and may account for the small
differences in the data, but it is striking how well the data match
between measurements taken over center-of-mass collision
energies that vary by more than an order of magnitude,
from $\sqrt{s}$ = 19.4 to 200 GeV.  If the same mechanisms
are responsible across this entire collision energy range,
then these mechanisms seem to have a weak dependence over the
interaction scale Q spanned by the world's data.

At forward rapidity $x_{F}$ is linearly proportional to the polarized parton 
momentum fraction $x_1$:
\begin{equation}
 x_F \equiv 2 p_L/\sqrt{s}\approx 2\langle z\rangle p_{\rm jet}/\sqrt{s}\approx\langle z\rangle x_1,
\end{equation}
where $\langle z \rangle$ is the mean momentum fraction of the hadron from the 
jet fragmentation. 
This suggests the possibility that these asymmetries are largely created by  some intrinsic 
function of $x$ that is only weakly dependent on the collision energy.

Further, from a comparison of the asymmetries of the pion isospin triplet at 
$\sqrt{s}=62.4$~GeV, see Fig.~\ref{fig:mpc62_phenix_brahms}, one can conclude
that the Sivers effect is not the dominant source of the observed transverse
asymmetries.
{\sc pythia} event generator studies show that the production of $\pi^{-}$ are almost 
equally from unfavored $u$ and favored $d$ quark fragmentation, while $\pi^{+}$ are 
almost exclusively from favored $u$ quark fragmentation.
At the same time, about three in four $\pi^{0}$ stem from $u$ quarks, with the other
fourth coming from $d$ quarks.  Because the Sivers effect comes from the initial
state quarks, the data can not be explained by these initial state effects alone,
under the assumption that the ratio of the $u$ and $d$ quark Sivers functions (especially
at high $x$) are the same as those extracted from SIDIS~\cite{Anselmino:2009ep},
According to these assumptions, one should naively expect a small Sivers effect
asymmetry for the $\pi^{+}$, which has roughly equivalent and canceling contributions
from $u$ and $d$ quarks. Instead a large asymmetry is observed for the $\pi^{+}$.

Collinear higher twist calculations predict that $A_{N}$ decreases with 
increasing transverse momentum once $p_{T}$ is of the same order as the 
partonic momentum scale $Q$ and both are much larger than 
$\Lambda_{QCD}$~\cite{Ji:2006ub}.
Where this turnover of the initially rising $A_{N}$ happens is largely 
unknown, though.
The cluster asymmetries in Fig.~\ref{fig:mpc200:AN_pt} have an extended 
$p_{T}$-range compared to previous measurements of $\pi^{0}$ mesons 
\cite{Abelev:2008}, but the data still do not allow for a conclusive answer 
for the onset of this drop of the asymmetry up to $p_{T}>4$~GeV/$c$.

The electromagnetic cluster contributions at $\sqrt{s}=200$~GeV are dominated 
by $\pi^{0}$ decays, as demonstrated in Fig.~\ref{fig:ptDecomp}.
With rising $p_{T}$, the fraction of direct and other photons increases
while the contribution from $\eta$ mesons does not change significantly.
A comparison of the cluster asymmetries with those of $\pi^{0}$ mesons from
STAR~\cite{Abelev:2008} in Fig.~\ref{fig:mpc200:AN_pt2} is largely consistent 
at small $x_{F}$ and statistically limited at $x_{F} > 0.47$, where the direct
photon contribution to the inclusive clusters becomes more important.
Transverse asymmetries of direct photons are of special interest in the future 
because they directly relate to the Sivers effect and its process dependence 
\cite{PhysRevLett.110.232301}.


The data presented in this paper provide crucial input to the 
long-standing question of the source of $A_{N}$ in hadronic collisions. The 
extended statistics of $A_{N}$ measurements for $\pi^{0}$ and $\eta$ at 
midrapidity, the cluster $A_{N}$ at 200~GeV, the complete isospin triplet of 
asymmetries at 62.4~GeV, and the extended range over beam collision 
energies, all quantitatively test the various theories seeking to explain 
these asymmetries.  In particular, the high statistics midrapidity data 
strongly constrain the presence of a gluon Sivers effect at midrapidity. 
The PHENIX data on \pizero transverse asymmetries, along with the world 
data, do not allow for a strong evolution with $Q^{2}$ in the combined 
effects from whatever causes these asymmetries. Finally, the mix of 
favored versus unfavored fragmentation for the three different pion 
states, and how these contribute to the asymmetries, also place 
constraints on the strengths of the contributing effects.



\section*{ACKNOWLEDGMENTS}   

We thank the staff of the Collider-Accelerator and Physics
Departments at Brookhaven National Laboratory and the staff of
the other PHENIX participating institutions for their vital
contributions.  We acknowledge support from the 
Office of Nuclear Physics in the
Office of Science of the Department of Energy,
the National Science Foundation, 
a sponsored research grant from Renaissance Technologies LLC, 
Abilene Christian University Research Council, 
Head of Department of Physics, University of Illinois at Urbana Champaign, 
Research Foundation of SUNY, and
Dean of the College of Arts and Sciences, Vanderbilt University 
(U.S.A),
Ministry of Education, Culture, Sports, Science, and Technology, 
the Japan Society for the Promotion of Science, and
Head Investigator, Graduate School of Science, Hiroshima University
(Japan),
Conselho Nacional de Desenvolvimento Cient\'{\i}fico e
Tecnol{\'o}gico and Funda\c c{\~a}o de Amparo {\`a} Pesquisa do
Estado de S{\~a}o Paulo (Brazil),
Natural Science Foundation of China (P.~R.~China),
Ministry of Education, Youth and Sports (Czech Republic),
Centre National de la Recherche Scientifique, Commissariat
{\`a} l'{\'E}nergie Atomique, and Institut National de Physique
Nucl{\'e}aire et de Physique des Particules (France),
Bundesministerium f\"ur Bildung und Forschung, Deutscher
Akademischer Austausch Dienst, and Alexander von Humboldt Stiftung (Germany),
Hungarian National Science Fund, OTKA (Hungary), 
Department of Atomic Energy and Department of Science and Technology (India),
Israel Science Foundation (Israel), 
National Research Foundation and WCU program of the 
Ministry Education Science and Technology (Korea),
Physics Department, Lahore University of Management Sciences (Pakistan),
Ministry of Education and Science, Russian Academy of Sciences,
Federal Agency of Atomic Energy, and 
Program Coordinator, Russian Research Center, Kurchatov Institute (Russia),
VR and Wallenberg Foundation (Sweden), 
the U.S. Civilian Research and Development Foundation for the
Independent States of the Former Soviet Union, 
the US-Hungarian Fulbright Foundation for Educational Exchange,
and the US-Israel Binational Science Foundation.

\section*{APPENDIX}

Data tables of measured $A_{N}$ with statistical and systematic 
uncertainties and cluster composition for cluster asymmetries at forward 
pseudorapidities.

\begin{table}[ht]
  \caption{Fractional composition of electromagnetic clusters in the MPC at $\sqrt{s}=200$~GeV for $x_F>0.4$, as shown in Fig.~\ref{fig:ptDecomp}.}
  \label{tbl:ClusterComposition}
  \begin{ruledtabular}\begin{tabular}{cccccccc}
      $\langle x_F \rangle$ & $\langle p_T \rangle$ & $\pi^0$ & $\eta$ & direct $\gamma$ & $h^{+,-}$ & other $\gamma$ \\
      \hline
      0.41 & 1.95 & 0.79 & 0.06 & 0.01 & 0.04 & 0.10  \\
      0.43 & 2.32 & 0.82 & 0.07 & 0.02 & 0.03 & 0.06  \\
      0.45 & 2.77 & 0.82 & 0.06 & 0.04 & 0.01 & 0.06  \\
      0.46 & 3.24 & 0.79 & 0.06 & 0.06 & 0.01 & 0.08  \\
      0.48 & 3.73 & 0.76 & 0.04 & 0.08 & 0.01 & 0.10  \\
      0.49 & 4.40 & 0.71 & 0.06 & 0.11 & 0.01 & 0.11  \\
      0.55 & 5.51 & 0.69 & 0.02 & 0.16 & 0.00 & 0.13  \\
  \end{tabular}\end{ruledtabular}
\end{table}

\begin{table*}[ht]
  \caption{The $A_{N}$ at $\sqrt{s}=62.4$~GeV as a function of $x_F$ for
  two pseudorapidity ranges, as shown in Fig.~\ref{fig:mpc62_xf}.}
  \label{tab:mpc62:AN_xf_etacuts}

  \begin{ruledtabular}\begin{tabular}{cccccc}
  & $\langle|x_{F}|\rangle$ & $\langle p_T \rangle$ & $A_{N}\pm\sigma_{\rm stat}\pm\sigma_{\rm syst}(x_F>0)$ & $A_{N}\pm\sigma_{\rm stat}\pm\sigma_{\rm syst}(x_F<0)$ \\
  \hline	
  $3.1<|\eta|<3.5$ &  0.25  &  0.41 & $-0.0123 \pm 0.0142 \pm 0.0037$ & $-0.0176 \pm 0.0141 \pm 0.0037$ \\
  $3.1<|\eta|<3.5$ &  0.35  &  0.59 & $ 0.0409 \pm 0.0117 \pm 0.0023$ & $ 0.0134 \pm 0.0116 \pm 0.0023$ \\
  $3.1<|\eta|<3.5$ &  0.44  &  0.74 & $ 0.0775 \pm 0.0149 \pm 0.0029$ & $-0.0204 \pm 0.0149 \pm 0.0029$ \\
  $3.1<|\eta|<3.5$ &  0.56  &  0.92 & $ 0.0843 \pm 0.0230 \pm 0.0066$ & $ 0.0057 \pm 0.0229 \pm 0.0066$ \\
  \\
  $3.5<|\eta|<3.8$ &  0.25  &  0.56 & $0.0273 \pm 0.0070 \pm 0.0018$ & $-0.0045 \pm 0.0070 \pm 0.0018$ \\
  $3.5<|\eta|<3.8$ &  0.35  &  0.77 & $0.0476 \pm 0.0082 \pm 0.0016$ & $-0.0129 \pm 0.0081 \pm 0.0016$ \\
  $3.5<|\eta|<3.8$ &  0.44  &  0.95 & $0.0465 \pm 0.0135 \pm 0.0026$ & $-0.0176 \pm 0.0134 \pm 0.0026$ \\
  $3.5<|\eta|<3.8$ &  0.54  &  1.15 & $0.0790 \pm 0.0304 \pm 0.0087$ & $-0.0025 \pm 0.0304 \pm 0.0087$ \\
  \end{tabular}\end{ruledtabular}
\end{table*}

\begin{table*}[ht]
  \caption{The $A_{N}$ as a function of $p_T$ at $\sqrt{s}=62.4$~GeV, as 
  shown in Fig.~\ref{fig:mpc62_pt}.}
  \label{tab:mpc62:AN_pt}

  \begin{ruledtabular}\begin{tabular}{cccccc}
  & $\langle p_{T}\rangle$ & $\langle |x_F| \rangle$ & $A_{N}\pm\sigma_{\rm stat}\pm\sigma_{\rm syst}(x_F>0)$ & $A_{N}\pm\sigma_{\rm stat}\pm\sigma_{\rm syst}(x_F<0)$ \\
  \hline	
  $3.1<\eta<3.8$ & 0.53   &  0.29 &  $0.0291 \pm 0.0120 \pm 0.0025$ &  $-0.0091 \pm 0.0117 \pm 0.0025$ \\
  $3.1<\eta<3.8$ & 0.67   &  0.34 &  $0.0577 \pm 0.0084 \pm 0.0019$ &  $-0.0055 \pm 0.0081 \pm 0.0018$ \\
  $3.1<\eta<3.8$ & 0.82   &  0.39 &  $0.0458 \pm 0.0093 \pm 0.0019$ &  $-0.0227 \pm 0.0093 \pm 0.0019$ \\
  $3.1<\eta<3.8$ & 1.01   &  0.45 &  $0.0687 \pm 0.0118 \pm 0.0014$ &  $-0.0112 \pm 0.0114 \pm 0.0013$ \\
  \end{tabular}\end{ruledtabular}
\end{table*}

\begin{table*}[ht]
  \caption{The $A_{N}$ at $\sqrt{s}=62.4$~GeV as a function of $x_F$, 
  as shown in Figs.~\ref{fig:mpc62_roots} and~\ref{fig:mpc62_phenix_brahms}.}
  \label{tab:mpc62:AN_xf}

  \begin{ruledtabular}\begin{tabular}{cccccc}
  & $\langle|x_{F}|\rangle$ & $\langle p_T \rangle$ & $A_{N}\pm\sigma_{\rm stat}\pm\sigma_{\rm syst}(x_F>0)$ & $A_{N}\pm\sigma_{\rm stat}\pm\sigma_{\rm syst}(x_F<0)$ \\
  \hline	
  $3.1<|\eta|<3.8$ & 0.25   &  0.52 &  $0.0193 \pm 0.0065 \pm 0.0017$ &  $-0.0067 \pm 0.0065 \pm 0.0017$ \\
  $3.1<|\eta|<3.8$ & 0.35   &  0.71 &  $0.0469 \pm 0.0067 \pm 0.0013$ &  $-0.0017 \pm 0.0066 \pm 0.0013$ \\
  $3.1<|\eta|<3.8$ & 0.44   &  0.86 &  $0.0605 \pm 0.0099 \pm 0.0019$ &  $-0.0182 \pm 0.0099 \pm 0.0019$ \\
  $3.1<|\eta|<3.8$ & 0.56   &  1.01 &  $0.0817 \pm 0.0182 \pm 0.0052$ &  $-0.0009 \pm 0.0181 \pm 0.0052$ \\
  \end{tabular}\end{ruledtabular}
\end{table*}

\begin{table*}[ht]
  \caption{The $A_{N}$ at $\sqrt{s}=200$~GeV as function of $p_{T}$ at 
   forward/backward rapidities in two different pseudorapidity ranges, 
   as shown in Fig.~\ref{fig:mpc200:AN_xf}.}
  \label{tbl:mpc200:AN_xf}
    
  \begin{ruledtabular}\begin{tabular}{cccrr}
    & $\langle|x_{F}|\rangle$ & $\langle p_{T}\rangle$ (GeV/$c$) & $A_{N}\pm\sigma_{\rm stat}\pm\sigma_{\rm syst}(x_{F}>0)$
                                                                 & $A_{N}\pm\sigma_{\rm stat}\pm\sigma_{\rm syst}(x_{F}<0)$ \\
    \hline
    $3.1<|\eta|<3.5$ & 0.28 & 2.1 & 0.0114 $\pm$ 0.0023 $\pm$ 0.0010 & -0.0016 $\pm$ 0.0023 $\pm$ 0.0010 \\
    $3.1<|\eta|<3.5$ & 0.32 & 2.4 & 0.0219 $\pm$ 0.0020 $\pm$ 0.0009 &  0.0022 $\pm$ 0.0020 $\pm$ 0.0009 \\
    $3.1<|\eta|<3.5$ & 0.37 & 2.7 & 0.0307 $\pm$ 0.0022 $\pm$ 0.0010 &  0.0016 $\pm$ 0.0023 $\pm$ 0.0010 \\
    $3.1<|\eta|<3.5$ & 0.43 & 3.1 & 0.0425 $\pm$ 0.0031 $\pm$ 0.0014 &  0.0010 $\pm$ 0.0030 $\pm$ 0.0014 \\
    $3.1<|\eta|<3.5$ & 0.50 & 3.6 & 0.0588 $\pm$ 0.0067 $\pm$ 0.0030 & -0.0016 $\pm$ 0.0065 $\pm$ 0.0029 \\
    $3.1<|\eta|<3.5$ & 0.60 & 4.3 & 0.0839 $\pm$ 0.0302 $\pm$ 0.0136 &  0.0480 $\pm$ 0.0261 $\pm$ 0.0117 \\
    \\
    $3.5<|\eta|<3.8$ & 0.28 & 1.5 & 0.0045 $\pm$ 0.0037 $\pm$ 0.0017 &  0.0015 $\pm$ 0.0038 $\pm$ 0.0017 \\
    $3.5<|\eta|<3.8$ & 0.33 & 1.8 & 0.0142 $\pm$ 0.0029 $\pm$ 0.0013 &  0.0042 $\pm$ 0.0029 $\pm$ 0.0013 \\
    $3.5<|\eta|<3.8$ & 0.37 & 2.0 & 0.0207 $\pm$ 0.0029 $\pm$ 0.0013 & -0.0010 $\pm$ 0.0028 $\pm$ 0.0013 \\
    $3.5<|\eta|<3.8$ & 0.43 & 2.3 & 0.0412 $\pm$ 0.0034 $\pm$ 0.0015 &  0.0026 $\pm$ 0.0033 $\pm$ 0.0015 \\
    $3.5<|\eta|<3.8$ & 0.50 & 2.7 & 0.0531 $\pm$ 0.0066 $\pm$ 0.0030 & -0.0015 $\pm$ 0.0064 $\pm$ 0.0029 \\
    $3.5<|\eta|<3.8$ & 0.60 & 3.2 & 0.0762 $\pm$ 0.0259 $\pm$ 0.0117 & -0.0149 $\pm$ 0.0239 $\pm$ 0.0108 \\
  \end{tabular}\end{ruledtabular}
\end{table*}

\begin{table*}[ht]
 \caption{The $A_{N}$ at $\sqrt{s}=200$~GeV in forward/backward 
rapidities ($|x_{F}|>0.4$), as shown in Fig. \ref{fig:mpc200:AN_pt}.
  }
  \label{tbl:mpc200:AN_pt}
    
  \begin{ruledtabular}\begin{tabular}{cccrr}
    & $\langle|x_{F}|\rangle$ & $p_{T}$ (GeV/$c$) & $A_{N}\pm\sigma_{\rm stat}\pm\sigma_{\rm syst}(x_{F}>0)$ 
                                                  & $A_{N}\pm\sigma_{\rm stat}\pm\sigma_{\rm syst}(x_{F}<0)$ \\
    \hline
    $3.1<|\eta|<3.8$ & 0.41 & 1.9 & 0.0154 $\pm$ 0.0155 $\pm$ 0.0070 &  0.0295 $\pm$ 0.0148 $\pm$ 0.0067 \\
    $3.1<|\eta|<3.8$ & 0.43 & 2.3 & 0.0394 $\pm$ 0.0037 $\pm$ 0.0017 &  0.0040 $\pm$ 0.0036 $\pm$ 0.0016 \\
    $3.1<|\eta|<3.8$ & 0.44 & 2.7 & 0.0458 $\pm$ 0.0034 $\pm$ 0.0015 & -0.0003 $\pm$ 0.0033 $\pm$ 0.0015 \\
    $3.1<|\eta|<3.8$ & 0.46 & 3.2 & 0.0542 $\pm$ 0.0045 $\pm$ 0.0020 & -0.0017 $\pm$ 0.0044 $\pm$ 0.0020 \\
    $3.1<|\eta|<3.8$ & 0.47 & 3.7 & 0.0487 $\pm$ 0.0071 $\pm$ 0.0032 & -0.0006 $\pm$ 0.0070 $\pm$ 0.0031 \\
    $3.1<|\eta|<3.8$ & 0.49 & 4.3 & 0.0524 $\pm$ 0.0112 $\pm$ 0.0050 &  0.0004 $\pm$ 0.0109 $\pm$ 0.0049 \\
    $3.1<|\eta|<3.8$ & 0.62 & 5.6 & 0.0192 $\pm$ 0.0261 $\pm$ 0.0117 &  0.0272 $\pm$ 0.0254 $\pm$ 0.0114 \\
    \end{tabular}\end{ruledtabular}
\end{table*}

\begin{table*}[ht]
  \caption{The $A_{N}$ as function of $p_{T}$ and $x_{F}$ at 
forward/backward rapidities, as shown in Fig.~\ref{fig:mpc200:AN_pt2}.}
  \label{tbl:mpc200:AN_pt2}
  \begin{ruledtabular}\begin{tabular}{ccrr}
    $\langle|x_{F}|\rangle$ & $\langle p_{T}\rangle$ (GeV/$c$) & $A_{N}\pm\sigma_{\rm stat}\pm\sigma_{\rm syst}(x_{F}>0)$ 
                                                               & $A_{N}\pm\sigma_{\rm stat}\pm\sigma_{\rm syst}(x_{F}<0)$ \\
    \hline
    0.27 & 1.3 & -0.0085 $\pm$ 0.0088 $\pm$ 0.0039 &  0.0176 $\pm$ 0.0088 $\pm$ 0.0040 \\
    0.28 & 1.7 &  0.0061 $\pm$ 0.0029 $\pm$ 0.0013 & -0.0011 $\pm$ 0.0030 $\pm$ 0.0013 \\
    0.28 & 2.1 &  0.0128 $\pm$ 0.0032 $\pm$ 0.0014 & -0.0029 $\pm$ 0.0032 $\pm$ 0.0014 \\
    0.28 & 2.6 &  0.0219 $\pm$ 0.0058 $\pm$ 0.0026 & -0.0064 $\pm$ 0.0057 $\pm$ 0.0026 \\
    \\		 	      		      	      	  		  
    0.32 & 1.6 &  0.0101 $\pm$ 0.0053 $\pm$ 0.0024 &  0.0037 $\pm$ 0.0053 $\pm$ 0.0024 \\
    0.32 & 2.0 &  0.0161 $\pm$ 0.0024 $\pm$ 0.0011 &  0.0052 $\pm$ 0.0025 $\pm$ 0.0011 \\
    0.33 & 2.4 &  0.0219 $\pm$ 0.0028 $\pm$ 0.0013 &  0.0026 $\pm$ 0.0029 $\pm$ 0.0013 \\
    0.33 & 3.0 &  0.0326 $\pm$ 0.0045 $\pm$ 0.0020 & -0.0042 $\pm$ 0.0045 $\pm$ 0.0020 \\
    \\		 	      		      	      	  		      
    0.37 & 1.8 &  0.0214 $\pm$ 0.0060 $\pm$ 0.0027 &  0.0032 $\pm$ 0.0058 $\pm$ 0.0026 \\
    0.37 & 2.2 &  0.0207 $\pm$ 0.0027 $\pm$ 0.0012 & -0.0024 $\pm$ 0.0027 $\pm$ 0.0012 \\
    0.37 & 2.7 &  0.0330 $\pm$ 0.0029 $\pm$ 0.0013 &  0.0018 $\pm$ 0.0030 $\pm$ 0.0013 \\
    0.38 & 3.2 &  0.0333 $\pm$ 0.0054 $\pm$ 0.0024 &  0.0096 $\pm$ 0.0054 $\pm$ 0.0024 \\
    0.38 & 3.9 &  0.0424 $\pm$ 0.0137 $\pm$ 0.0062 & -0.0041 $\pm$ 0.0135 $\pm$ 0.0061 \\
    \\		 	      		      	      	  		      
    0.42 & 2.1 &  0.0333 $\pm$ 0.0063 $\pm$ 0.0029 &  0.0020 $\pm$ 0.0061 $\pm$ 0.0027 \\
    0.43 & 2.4 &  0.0368 $\pm$ 0.0036 $\pm$ 0.0016 &  0.0003 $\pm$ 0.0035 $\pm$ 0.0016 \\
    0.43 & 3.0 &  0.0499 $\pm$ 0.0041 $\pm$ 0.0018 &  0.0038 $\pm$ 0.0040 $\pm$ 0.0018 \\
    0.43 & 3.5 &  0.0458 $\pm$ 0.0073 $\pm$ 0.0033 &  0.0012 $\pm$ 0.0071 $\pm$ 0.0032 \\
    0.44 & 4.1 &  0.0459 $\pm$ 0.0125 $\pm$ 0.0056 & -0.0057 $\pm$ 0.0122 $\pm$ 0.0055 \\
    \\		 	      		      	      	  		      
    0.48 & 2.2 &  0.0463 $\pm$ 0.0350 $\pm$ 0.0158 &  0.0723 $\pm$ 0.0338 $\pm$ 0.0152 \\
    0.50 & 2.6 &  0.0551 $\pm$ 0.0075 $\pm$ 0.0034 & -0.0058 $\pm$ 0.0073 $\pm$ 0.0033 \\
    0.50 & 3.1 &  0.0589 $\pm$ 0.0084 $\pm$ 0.0038 & -0.0043 $\pm$ 0.0082 $\pm$ 0.0037 \\
    0.50 & 3.7 &  0.0660 $\pm$ 0.0113 $\pm$ 0.0051 &  0.0080 $\pm$ 0.0109 $\pm$ 0.0049 \\
    0.51 & 4.4 &  0.0339 $\pm$ 0.0186 $\pm$ 0.0084 & -0.0023 $\pm$ 0.0176 $\pm$ 0.0079 \\
    \\		 	      		      	      	  		      
    0.58 & 3.0 &  0.0724 $\pm$ 0.0270 $\pm$ 0.0122 & -0.0288 $\pm$ 0.0261 $\pm$ 0.0117 \\
    0.60 & 3.7 &  0.0772 $\pm$ 0.0221 $\pm$ 0.0100 & -0.0044 $\pm$ 0.0212 $\pm$ 0.0095 \\
    0.61 & 4.7 &  0.0632 $\pm$ 0.0389 $\pm$ 0.0175 &  0.0453 $\pm$ 0.0375 $\pm$ 0.0169 \\
  \end{tabular}\end{ruledtabular}
\end{table*}

\begin{table*}[ht]
  \caption{The $A_{N}$ of $\pi^{0}$ mesons at $\sqrt{s}=200$~GeV at midrapidity 
as function of $p_{T}$, as shown in Fig.~\ref{fig:emc200}. The data in 
slightly forward and backward kinematics ($0.2<|\eta|<0.35$) are subsets 
of the full data set ($|\eta|<0.35$).
  }
  \label{tbl:emc200_pi0}

  \begin{ruledtabular}\begin{tabular}{crrr}
      $\langle p_{T}\rangle$ (GeV/$c$) & $A_{N}\pm\sigma_{\rm stat}\pm\sigma_{\rm syst}$
                                       & $A_{N}\pm\sigma_{\rm stat}\pm\sigma_{\rm syst}$
				       & $A_{N}\pm\sigma_{\rm stat}\pm\sigma_{\rm syst}$\\
                                       & $(|\eta|<0.35)$
                                       & $(0.2<|\eta|<0.35,x_{F}>0)$
				       & $(0.2<|\eta|<0.35,x_{F}<0)$ \\
      \hline
       1.5 &  0.0008 $\pm$ 0.0006 $\pm$ 0.0002 &  0.0012 $\pm$ 0.0014 $\pm$ 0.0003 &  0.0020 $\pm$ 0.0014 $\pm$ 0.0003 \\
       2.4 &  0.0006 $\pm$ 0.0006 $\pm$ 0.0002 &  0.0021 $\pm$ 0.0013 $\pm$ 0.0003 &  0.0012 $\pm$ 0.0013 $\pm$ 0.0003 \\
       3.4 &  0.0002 $\pm$ 0.0011 $\pm$ 0.0003 &  0.0025 $\pm$ 0.0025 $\pm$ 0.0005 & -0.0009 $\pm$ 0.0025 $\pm$ 0.0005 \\
       4.4 &  0.0013 $\pm$ 0.0022 $\pm$ 0.0006 &  0.0030 $\pm$ 0.0053 $\pm$ 0.0011 & -0.0016 $\pm$ 0.0053 $\pm$ 0.0011 \\
       5.4 &  0.0004 $\pm$ 0.0045 $\pm$ 0.0009 &  0.0139 $\pm$ 0.0106 $\pm$ 0.0021 & -0.0072 $\pm$ 0.0106 $\pm$ 0.0021 \\
       6.4 & -0.0071 $\pm$ 0.0082 $\pm$ 0.0016 & -0.0368 $\pm$ 0.0197 $\pm$ 0.0039 & -0.0086 $\pm$ 0.0198 $\pm$ 0.0040 \\
       7.4 & -0.0062 $\pm$ 0.0136 $\pm$ 0.0027 & -0.0699 $\pm$ 0.0337 $\pm$ 0.0067 &  0.0587 $\pm$ 0.0337 $\pm$ 0.0067 \\
       8.4 &  0.0036 $\pm$ 0.0210 $\pm$ 0.0052 &  0.0116 $\pm$ 0.0801 $\pm$ 0.0160 & -0.0026 $\pm$ 0.0935 $\pm$ 0.0187 \\
       9.4 &  0.0059 $\pm$ 0.0318 $\pm$ 0.0064 &      -                         &      -                         \\
      10.8 &  0.0331 $\pm$ 0.0387 $\pm$ 0.0077 &      -                         &      -                         \\
  \end{tabular}\end{ruledtabular}
\end{table*}

\begin{table*}[ht]
  \caption{The $A_{N}$ of $\eta$ mesons at $\sqrt{s}=200$~GeV at 
midrapidity as function of $p_{T}$, as shown in Fig.~\ref{fig:emc200}. The 
data in slightly forward and backward kinematics ($0.2<|\eta|<0.35$) are 
subsets of the full data set ($|\eta|<0.35$).
  }
  \label{tbl:emc200_eta}
    
  \begin{ruledtabular}\begin{tabular}{cccr}
      $\langle p_{T}\rangle$ (GeV/$c$) & $A_{N}\pm\sigma_{\rm stat}\pm\sigma_{\rm syst}$
                                       & $A_{N}\pm\sigma_{\rm stat}\pm\sigma_{\rm syst}$
				       & $A_{N}\pm\sigma_{\rm stat}\pm\sigma_{\rm syst}$\\
                                       & $(|\eta|<0.35)$
                                       & $(0.2<|\eta|<0.35,x_{F}>0)$
				       & $(0.2<|\eta|<0.35,x_{F}<0)$ \\
      \hline
      2.4 & -0.0069 $\pm$ 0.0049 $\pm$ 0.0010 &  -0.0169 $\pm$ 0.0125 $\pm$ 0.0025 &  0.0070 $\pm$ 0.0126 $\pm$ 0.0025 \\
      3.4 & -0.0024 $\pm$ 0.0057 $\pm$ 0.0012 &  -0.0355 $\pm$ 0.0154 $\pm$ 0.0031 &  0.0040 $\pm$ 0.0155 $\pm$ 0.0031 \\
      4.4 & -0.0019 $\pm$ 0.0099 $\pm$ 0.0020 &  -0.0073 $\pm$ 0.0265 $\pm$ 0.0053 & -0.0336 $\pm$ 0.0265 $\pm$ 0.0053 \\
      5.4 &  0.0292 $\pm$ 0.0171 $\pm$ 0.0034 &   0.0178 $\pm$ 0.0452 $\pm$ 0.0090 & -0.0327 $\pm$ 0.0453 $\pm$ 0.0091 \\
      6.4 & -0.0458 $\pm$ 0.0285 $\pm$ 0.0057 &   0.0021 $\pm$ 0.0987 $\pm$ 0.0197 &  0.0896 $\pm$ 0.1131 $\pm$ 0.0226 \\
      7.4 &  0.0035 $\pm$ 0.0431 $\pm$ 0.0086 &       -                         &       -                        \\
      9.1 &  0.0842 $\pm$ 0.0550 $\pm$ 0.0110 &       -                         &       -                        \\
  \end{tabular}\end{ruledtabular}
\end{table*}

\clearpage



\end{document}